\documentclass[11pt]{article}
\usepackage{vietnam,epsfig}

\def\Journal#1#2#3#4{{#1} {\bf #2}, #3 (#4)}

\def\PLB{{\em Phys. Lett.}  B}
\def\PRL{\em Phys. Rev. Lett.}

\def\ZPC{{\em Z. Phys.} C}
\def\EPJ{{\em Eur. Phys. J.} C}
\def\IJMPA{{\em Int. J. Mod. Phys.} A} 

\def\be{\begin{equation}}
\def\ee{\end{equation}}
\def\bea{\begin{eqnarray}}
\def\eea{\end{eqnarray}}

\begin{document}
\vspace*{4cm}
\title{SEARCHES FOR EXOTICA AT LEP}

\author{ F.-L. NAVARRIA }

\address{Dip. di Fisica, Univ. di Bologna, V.le Berti-Pichat 6/2,
I-40127 Bologna, Italy \\
INFN, Sezione di Bologna, V.le Berti-Pichat 6/2, 
I-40127 Bologna, Italy}

\maketitle\abstracts{
The results of various searches for new physical phenomena beyond the 
Standard Model using data from the four LEP experiments are 
summarized. Topics presented include the search for flavour-changing 
neutral currents with single top production, compositeness leading to
the production of excited leptons, and manifestations of  
extra dimensions.}

\section{Introduction}
The Standard Model (SM) works extremely well so far, yet it is believed 
to be only a low-energy effective theory and new physics beyond the SM
is expected to appear. 
Some of the motivations for looking beyond the SM are neutrino oscillations, 
the problems related to the unification of forces (hierarchy problem), 
and the origin of the mass hierarchy. 
Theoretical prejudices vary a lot, but indicate, in some cases, that the 
new scale could be as low as 1 TeV. New phenomena could therefore have 
been accessible with the Large Electron Positron 
Collider at CERN. LEP was stopped at the end of 2000, after reaching 
$\sqrt{s}$ = 209 GeV, and will be replaced by the Large Hadron Collider 
(LHC) in 2007. The four LEP experiments, ALEPH, DELPHI, L3, and OPAL, 
%~\cite{{ale}{del}{l3}{opa}}
have collected almost 2.4 fb$^{-1}$ of data at the highest LEP2 energies 
(above 
182 GeV).  Here several searches for new exotic phenomena will be 
presented, namely the searches for single top production in Section 2, 
excited leptons in Section 3, and extra dimensions in Section 4. Many of 
the final results have been available since some time, but in a few cases 
the experiments have not yet produced the analysis of the full data sets 
(in particular the higher energy data) because of lack of manpower, or 
they are just finalizing the results.  
A LEP working group was set up to combine the results of the four 
experiments and obtain the best limits in the search for exotic 
phenomena. Two combinations were performed in 
2001 often using preliminary data~\cite{wg}.  
A combination of the limits on extra dimensions has 
been produced recently~\cite{wg}. 

\section{Single top production}
\begin{figure}[ht]
\centerline{\hbox{
\psfig{figure=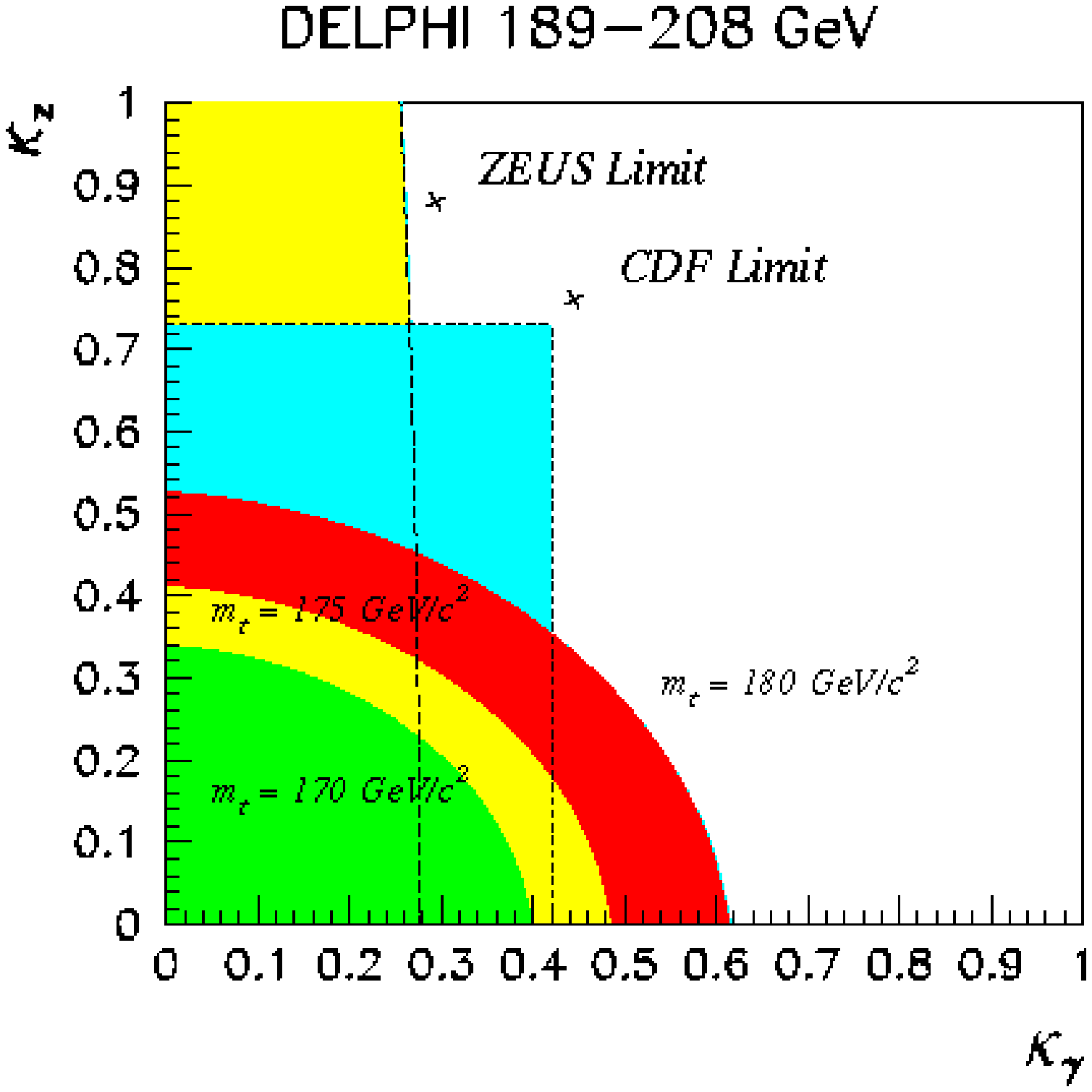,height=2.1in}
\epsfxsize=5.4cm\epsffile{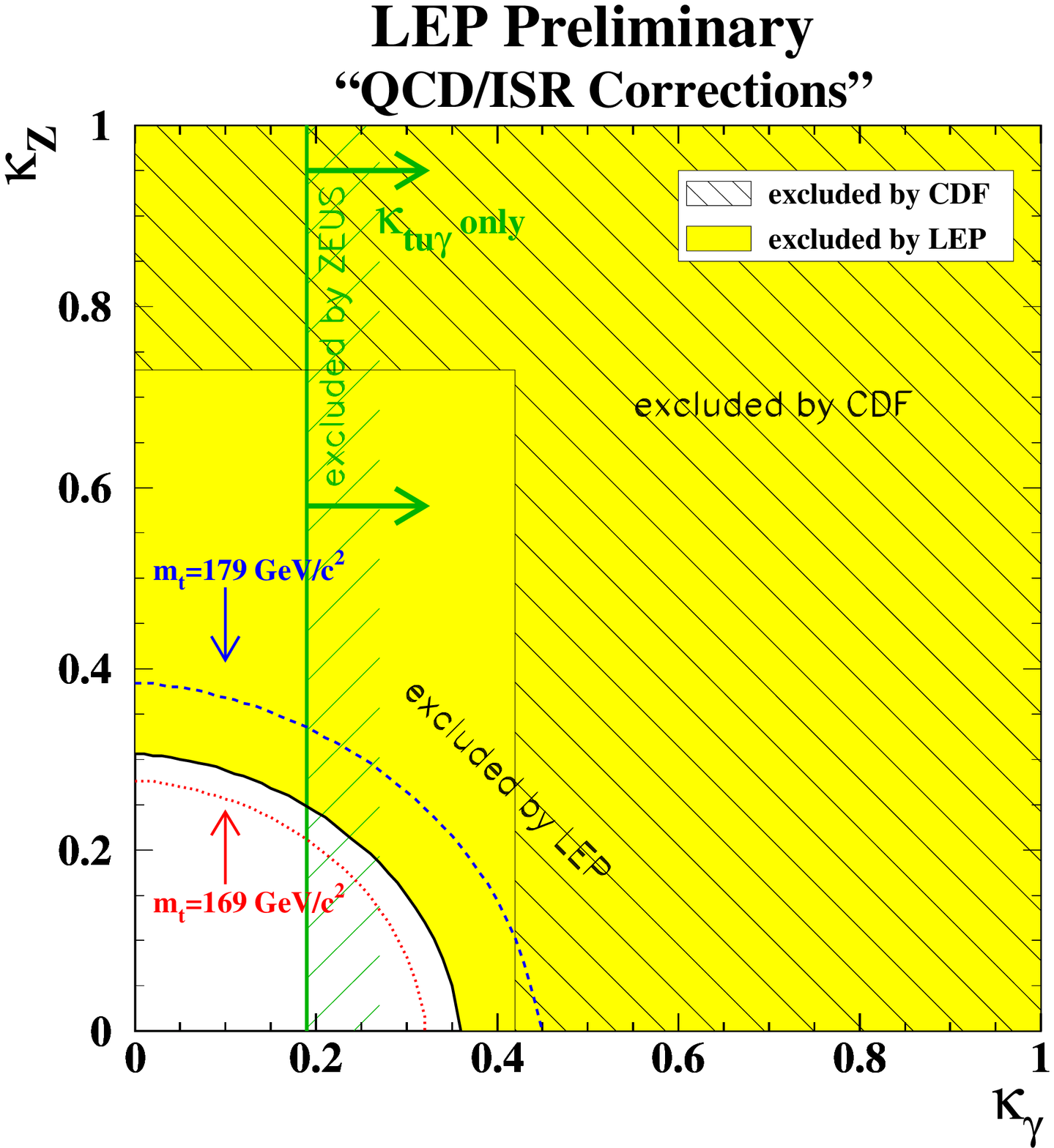}
\epsfxsize=5.4cm\epsffile{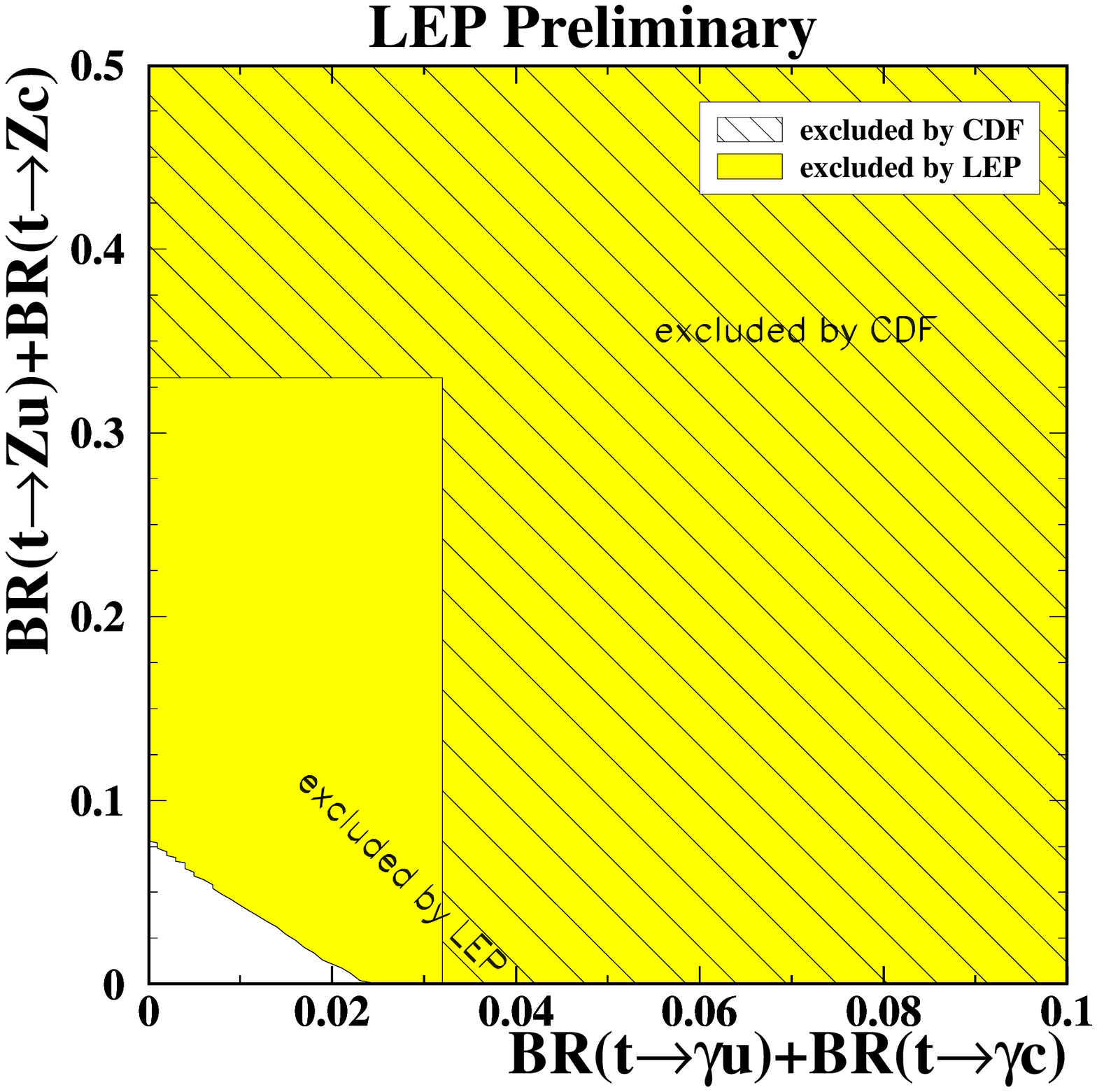}
           }     }
\caption{Left: Search for single top production, 
95$\%$ CL area excluded by DELPHI in the $\kappa_\gamma$, $\kappa_Z$ plane. 
Middle and right: anomalous couplings and top BRs excluded by combined LEP 
data and by CDF.  
\label{fig:st-lep}}
\end{figure}

Flavor Changing Neutral Currents (FCNC) are absent in the SM at the tree 
level and severely suppressed even at the one-loop level. For example, at 
LEP2 energies the production of a single top quark, e$^+$e$^- \rightarrow$ 
t$\bar{\rm c}$ + c.c. (t$\bar{\rm u}$ + c.c.),
%\footnote{Here and in the following, 
%when relevant, charge conjugate final are also considered.} 
is present at the one-loop level in the SM, 
but the cross-section is expected to be only $O(10^{-9}$ fb).
%~\cite{huang} 
The SM process is therefore 
totally invisible and single t production can be used as a probe 
for new physics. FCNC can also be searched for at HERA via single t  
production, ep $\rightarrow$ etX, and at the Tevatron via rare top decay,
t $\rightarrow$ Z($\gamma$) c(u), since the one-loop level
BR[t $\rightarrow$ (Z,g,$\gamma$) + c(u)] is predicted to be 
$<$ $10^{-10}$ in the SM. 
%~\cite{eilam}
Several extensions of the SM predict enhancements 
of the single t production cross-sections or larger BRs. 
%Supersymmetry and multiple Higgs 
%doublet models predict FCNC at the tree level. FCNC coupled singlet quarks,
%compositness, or dynamical ew symmetry breaking~\cite{fcnc-th} predict BRs 
%up to $\sim$ $10^{-2}$.    
FCNC t production and decay can be expressed in terms of 
anomalous couplings $\kappa_\gamma$ and  $\kappa_Z$,~\cite{obratzsov}
with a scale assumed to be equal to m$_t$. Events of the type 
e$^+$e$^- \rightarrow$ t[$\rightarrow$ b W($\rightarrow$ q$\bar{\rm q}$, 
$l\nu_l$)] $\bar{\rm c}$ or $\bar{\rm u}$) have been searched for using 
b-tag and kinematic 
variables. No excess was found over the SM background and 
exclusion limits were derived for the production cross-sections. 
%FCNC t production and decay can be expressed in terms of anomalous 
%couplings $\kappa_\gamma$ and  $\kappa_Z$, so that the limits can be 
%traslated into an excluded area in the $\kappa_\gamma$, $\kappa_Z$ plane. 
All experiments have now produced final 
results~\cite{{st-al},{st-de},{st-l3},{st-op}}.
%see Table~\ref{tab:st-exp}, where ALEPH, OPAL (DELPHI, L3) use m$_t$ = 
%174 (175) GeV. 
%\begin{table}[t]
%\caption{95\% CL limits on anomalous couplings and BRs from the search 
%for single t production.\label{tab:st-exp}}
%\vspace{0.4cm}
%\begin{center}
%\begin{tabular}{|c|c|c|c|l|}
%\hline
%& $\kappa_\gamma$($\kappa_Z$=0)  & $\kappa_Z$($\kappa_\gamma$=0) & $BR_\gamma$ & $BR_Z$ \\  
%\hline
%ALEPH  & (0.49) & 0.42 & (0.042) & 0.140 \\ 
%DELPHI & 0.486 & 0.411 & & \\
%L3     & 0.43 & 0.37 & 0.041 & 0.137 \\
%OPAL   & 0.48 & 0.41 & & 
%\\ \hline
%\end{tabular}
%\end{center}
%\end{table}
Figure~\ref{fig:st-lep} (left) shows as an example the limits published 
recently by DELPHI~\cite{st-de}:
also shown are the limits obtained by ZEUS 
%~\cite{st-zeus}
at HERA\footnote{The experiments at HERA are mostly sensitive to 
$\kappa_{tu\gamma}$, since Z exchange is suppressed owing to the large Z 
mass and because t is produced at large $x$. 
The difference in the ZEUS exclusion limit in Fig.~\ref{fig:st-lep} left and 
middle is due to a factor of $\sqrt{2}$ difference in the Lagrangian used 
at LEP and HERA: the correct one, from the LEP point of view, is the one in 
Fig.~\ref{fig:st-lep} left.}, and by CDF in Run I at the 
Tevatron. 
%~\cite{st-cdf} 
H1 
%~\cite{st-h1}
has produced limits which are looser than those of ZEUS because it has 
observed an excess of leptonic events at high p$_T$. However, this excess, 
is not observed in other channels (hadronic) or by ZEUS. 
The limits on the anomalous couplings and on the BRs derived from 
the combined LEP data are shown in Fig.~\ref{fig:st-lep} (middle) and 
Fig.~\ref{fig:st-lep} (right): a BR[t$\rightarrow$Zc(u)] $>$ 0.081 is 
excluded at 95\% CL for $\kappa_\gamma$ = 0.~\cite{wg}

\vspace*{-0.1cm}
\section{Excited leptons}

\begin{figure}[ht]
\centerline{\hbox{
\epsfxsize=7.cm\epsffile{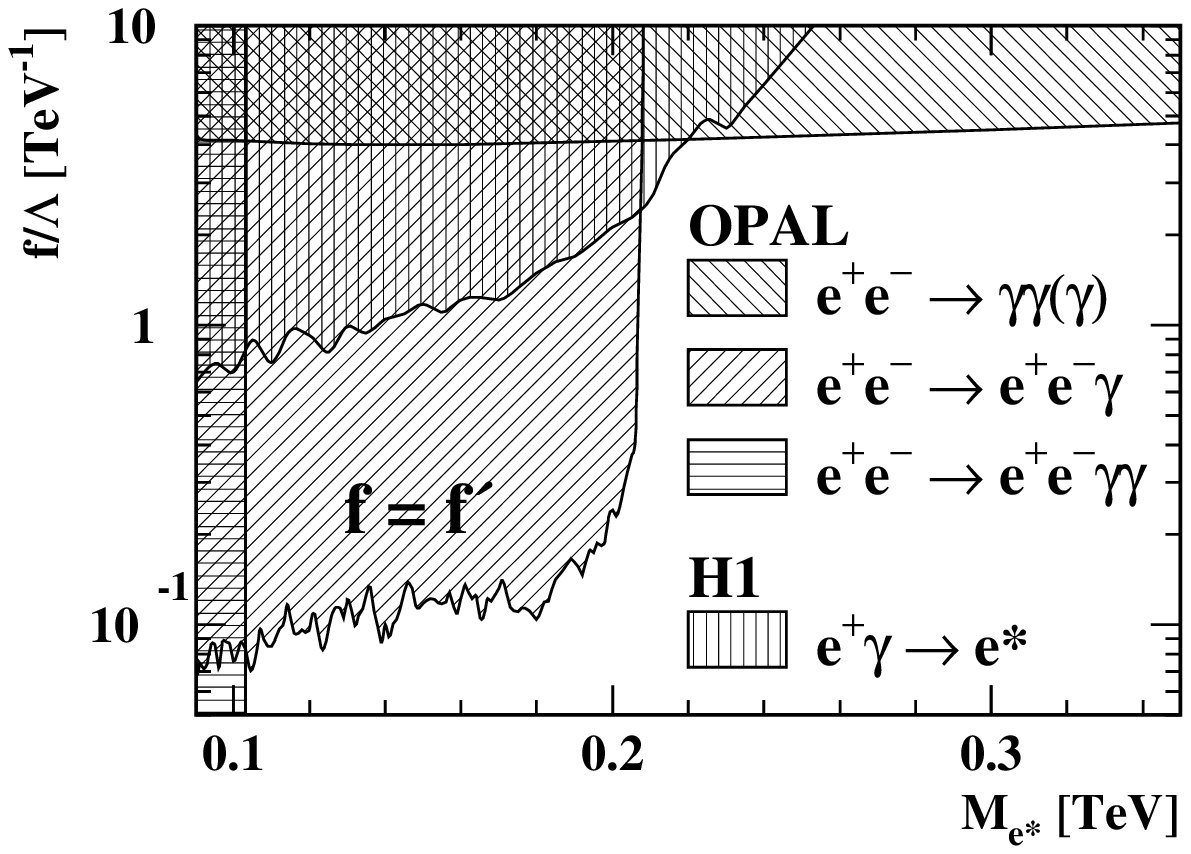}
\psfig{figure=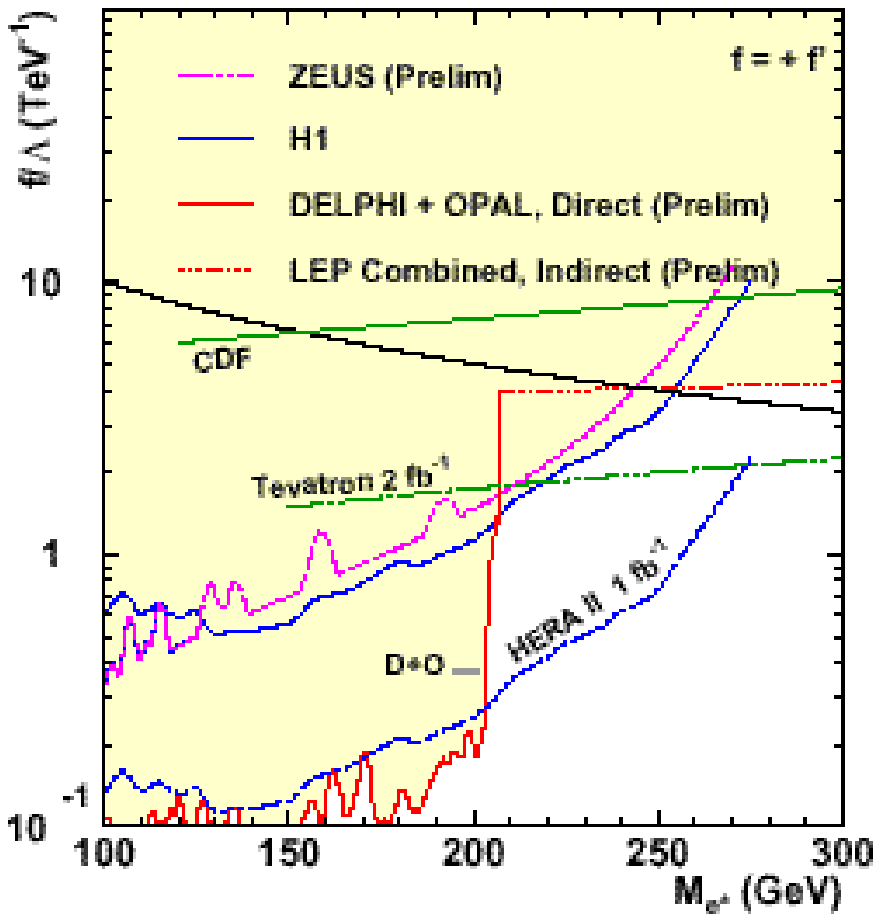,height=2.0in}
           }     }
\caption{Left: 95$\%$ CL area excluded by 
%combined DELPHI and 
OPAL data 
in the search for 
%single excited lepton production. 
e$^*$: the limits from pair and single production, 
and the indirect limit are shown separately. The limits obtained by H1 
at HERA are also shown.  
Right: DELPHI and OPAL combined search for 
e$^*$; the indirect limit from 
e$^+$e$^-$$\rightarrow$$\gamma$$\gamma$ results  
from the combination of the four LEP experiments. 
The decreasing curve shows the hyperbola f/$\Lambda$ = 1/m$_{e^*}$: 
m$_{e^*}$ $>$ 248 GeV, for f = 1, $\Lambda$ = m$_{e^*}$. 
Limits and sensitivities of HERA and Tevatron 
experiments are also shown. 
\label{fig:ex-do}}
\end{figure}

The existence of three families of quarks and leptons is a 
strong motivation to look for substructure. In composite 
models~\cite{exl-th}, quarks, 
leptons and gauge bosons are composite with an associated energy scale,
$\Lambda$. At s $<<\Lambda^2$ there could be manifestations of these 
subconstituents through anomalous decay modes, anomalous electric and 
magnetic multipoles, excited fermions, leptoquarks and contact terms.
For example, excited leptons would decay promptly emitting a gauge boson,
$\gamma$, W or Z, and an ordinary lepton of the same family.
The possible processes are $l^*$ $\rightarrow$ 
$l\gamma$, $\nu$W, $l$Z; $\nu^*$ $\rightarrow$ $\nu\gamma$, $l$W, $\nu$Z.
BR, topologies and efficiencies depend on the relative strengths, f 
and f$^\prime$, of the weights multiplying the SM couplings.
For instance, if f = f$^\prime$, 
$\nu^*$ $\rightarrow$ $\nu$$\gamma$ is forbidden; if f = -f$^\prime$,
$l^*$ $\rightarrow$ $l\gamma$ is forbidden. At LEP two classes of 
processes were 
investigated: pair production, e$^+$e$^-$ $\rightarrow$ $l^*$$l^*$, 
$\nu^*$$\nu^*$, with a discovery limit m$_*$$\sim$$\sqrt{s}$/2; single 
production, e$^+$e$^-$ $\rightarrow$ $l^*$$l$, $\nu^*$$\nu$, with a
discovery limit m$_*$$\sim$$\sqrt{s}$. The single production 
cross-sections depend on the ratios f,f$^\prime$/$\Lambda$ and on m$_{l^*}$. 
For excited electrons 
a limit beyond $\sqrt{s}$ can also be obtained by looking at the reaction
e$^+$e$^-$ $\rightarrow$ $\gamma$$\gamma$, which is sensitive to 
virtual e$^*$ exchange in the t-channel in addition to the SM 
contribution.~\cite{ggwg} L3~\cite{exl-l3} and OPAL~\cite{exl-op} 
have published results 
up to the highest energies, while DELPHI results are still 
preliminary~\cite{exl-de}. A preliminary combination of DELPHI and 
OPAL data was performed in 2001~\cite{wg}, and a new one should be 
available soon. No excess above the SM background was observed and 
upper limits for the single production cross sections and the f,f$^\prime$ 
couplings divided by the compositess scale were 
derived assuming a model with an excited doublet with L,R 
components (pair production allows excited leptons with masses below 
%about 
95-103 GeV to be excluded, for any coupling). Figure~\ref{fig:ex-do} 
(left) shows the 
%preliminary 
limits 
obtained by 
%DELPHI and 
OPAL for f/$\Lambda$ vs 
%m$_{l^*}$ 
m$_{e^*}$ assuming f = f$^\prime$.~\cite{exl-op}  
%The wavy exclusion lines are due to the presence of candidates smeared by 
%the resolution. 
Figure~\ref{fig:ex-do} (right) shows the preliminary limits 
obtained at LEP (directly by DELPHI and OPAL, indirectly by all four 
experiments) for the excited electron assuming f = f$^\prime$~\cite{wg}: they are 
compared with the limits obtained at HERA
and at the Tevatron~\cite{perez}. For masses below 200 GeV they 
will remain the best limits for some time, until HERA exceeds an 
integrated luminosity of $\sim$ 1 fb$^{-1}$. In addition, HERA can search 
only for e$^*$  and $\nu_e^*$.
%~\cite{exl-hera} 
     
\vspace*{-0.1cm}
\section{Search for extra dimensions}

As m$_{ew}$ $<<$ m$_{Pl}$, unification of gravity with other forces 
could only take place at extremely high energies, well out of the reach 
of particle accelerators. Various solutions of the hierarchy problem 
have been proposed, which involve the existence of extra spatial 
dimensions (large ED, warped ED, etc.). These solutions can be tested 
at present accelerators and at the LHC.

\subsection{Large extra dimensions}

\begin{figure}[ht]
\centerline{\hbox{
\epsfxsize=6.3cm\epsffile{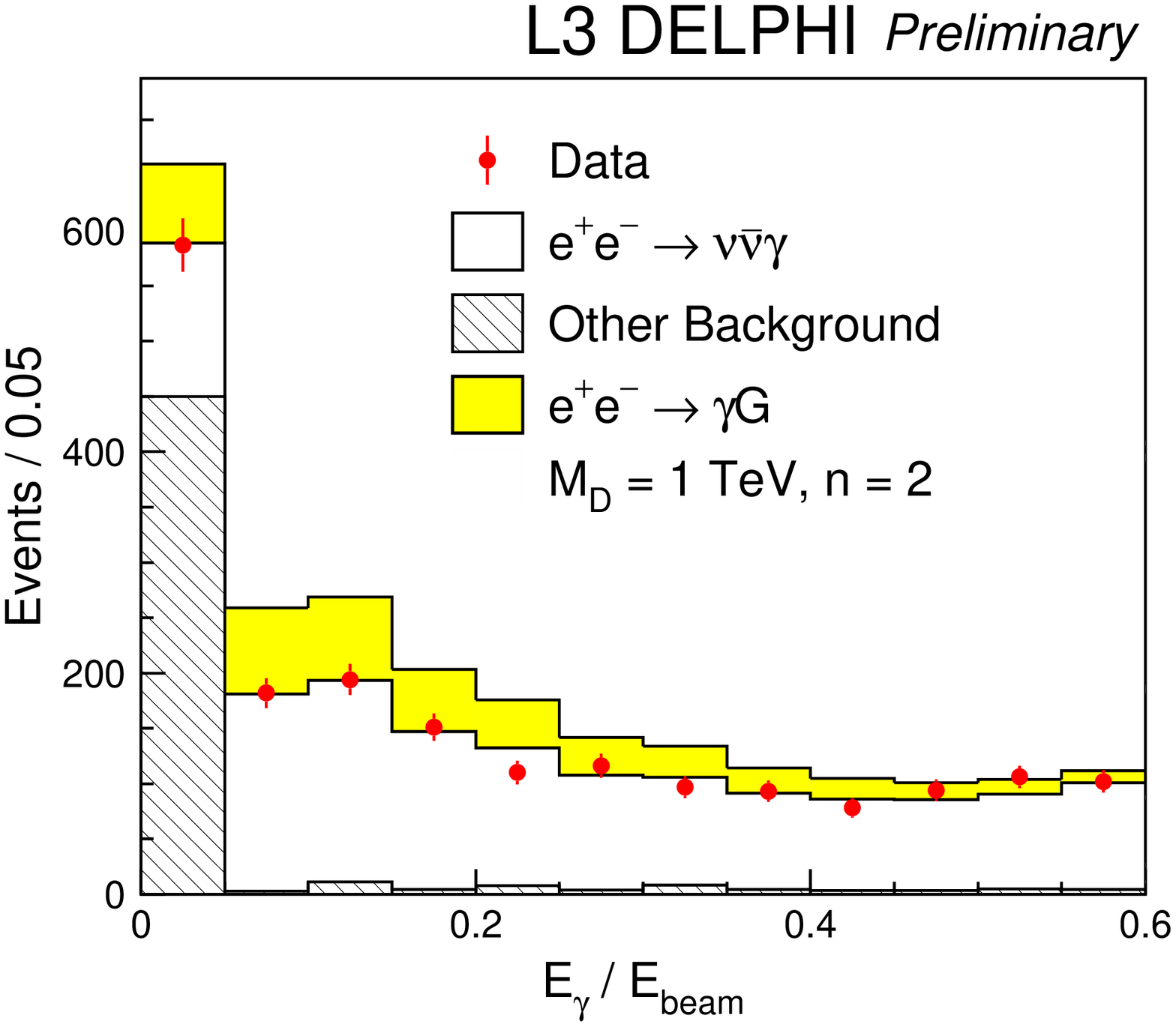}
\epsfxsize=5.4cm\epsffile{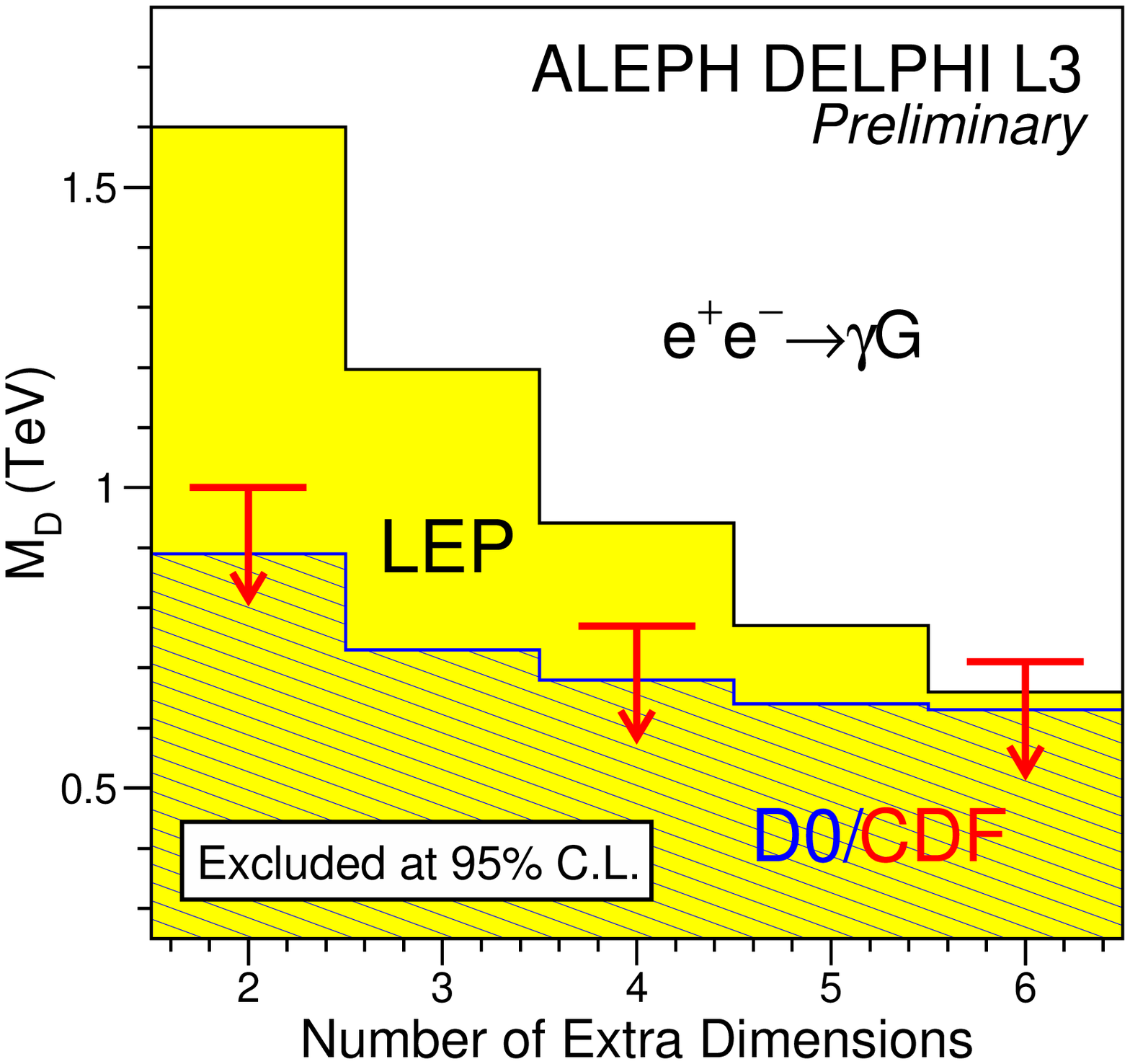}
           }     }
\caption{Left: photon energy spectrum in single $\gamma$ events. 
Right: the limits on M$_D$ as a function of the number n of extra dimensions. 
\label{fig:ex-dim}}
\end{figure}

The ADD model~\cite{add} assumes a D dimensional space-time, with 
D = 3+n+1, n being the extra space dimensions: SM fields live in a 
3-dimensional rigid brane, whilst gravity is allowed propagate in 
the bulk, thus becoming diluted in the big extra space. 
If the n extra dimensions are compactified on a torus (flat ED), one has
m$_{Pl}^2$ $\sim$ R$^n$M$_D^{n+2}$, with M$_D$ the D-dimensional Planck 
mass, and if M$_D$ is $\sim$ 1 TeV, thus eliminating the hierarchy 
problem, the compactification radius would be 
R = 0.3 mm, 10 pm, 30 fm for n-dim = 2, 4, 6, respectively. 
The bulk graviton is expanded in a Kaluza-Klein tower of massive states, 
with $\Delta$m $\sim$ 1/R, forming almost a continuous spectrum and being 
weakly coupled. At LEP one can look for i) graviton emission in 
e$^+$e$^-$ $\rightarrow$ $\gamma$G, i.e. $\gamma$+missing energy, 
ii) graviton exchange in boson and fermion pair production, 
e$^+$e$^-$ $\rightarrow$ $\gamma$$\gamma$, WW, ZZ, f$\bar{\rm f}$. 
Experimentally, n=1 is excluded by the behaviour of Newton's law at 
solar system scales. Contrary to ew and strong forces, which have been
tested down to $\sim$ (100 GeV/$c$)$^{-1}$ $\sim$ 10$^{-15}$ mm, gravity
so far has been tested down only to the 0.1 mm scale, whilst improved 
experiments could decrease this limit by an order of magnitude.
%~\cite{slac-gr}
Limits on ED for n = 2 can also be derived from astrophysical and cosmological 
arguments~\cite{pdg}, but tend to be less constraining for larger n. 

i) Graviton emission. The angular distribution for  e$^+$e$^-$ 
$\rightarrow$ $\gamma$G is peaked at small E$_\gamma$ and $\theta_\gamma$, 
and one of the principal SM backgrounds is e$^+$e$^-$ $\rightarrow$ 
$\nu$$\bar{\nu}$$\gamma$. The combination of the DELPHI and L3 
single $\gamma$ energy spectrum is shown in Fig.~\ref{fig:ex-dim} (left). 
No deviation is observed with respect to the SM background. 
The exclusion limits for M$_D$ as a function of n are presented in 
Fig.~\ref{fig:ex-dim} (right): the ALEPH data have been included 
reconstructing the likelihoods from the fitted M$_D$ vs n. OPAL data at the 
highest energies have not been yet analysed. For n = 2, 3, 4, 5, 6 one has 
M$_D$$>$1.60, 1.20, 0.94, 0.77, 0.66 TeV.~\cite{wg} The limits from CDF and D0 
at the Tevatron are also shown: they equal the LEP limits only at large n.    
     
ii) Graviton exchange. The most sensitive channel is Bhabha scattering.
The scale is not the same as M$_D$ but one has to introduce a cut-off, 
M$_S$, with M$_D^4$ = (2/$\pi\lambda$) M$_S^4$, $|$$\lambda$$|$ $\sim$ 1. 
Three terms contribute to the angular distribution, the SM term, G 
exchange and the interference term ($\sim$ $\lambda$/M$_S^4$). No deviation 
from the SM predictions 
is observed and typical limits are M$_S$ $>$ 1.20(1.09) TeV for 
$\lambda$ = +1(-1) at 95\% CL combining the data from the four 
experiments.~\cite{ewwg} Similarly, the combined LEP results 
from e$^+$e$^-$ $\rightarrow$ $\gamma$$\gamma$ are M$_S$ $>$ 0.93(1.01) 
TeV for $\lambda$ = +1(-1).~\cite{ggwg}   

\subsection{Search for branons}

If the brane is permitted to vibrate (instead of being rigid as in the ADD 
model), there will be brane fluctuations along the ED, and associated new  
pseudoscalar particles (branons, $\tilde{\pi}$) will appear. Branons could 
be dark matter candidates in low-tension brane worlds~\cite{dobado}. The 
search at LEP was performed by L3,~\cite{l3-br} complementing the search 
for G emission. In fact, if f (the brane tension) $>>$ M$_F$ (the gravity 
scale), then G is accessible first; if f $\ll$ M$_F$, then one should look 
for $\tilde{\pi}$ produced in pairs, e$^+$e$^-$ $\rightarrow$ 
$\tilde{\pi}$$\tilde{\pi}$$\gamma$ or $\tilde{\pi}$$\tilde{\pi}$Z, 
looking at events with photons/jets and missing energy. Since no excess 
is observed above the SM background, L3 sets a limit M $>$ 103 GeV 
at 95\% CL on the branon mass for an elastic brane 
(f$\rightarrow$0), and f $>$ 180 GeV (M=0)~\cite{l3-br}.   
   
\subsection{Search for the radion}

In the Randall-Sundrum (RS) model~\cite{rs} there are two branes: the SM is 
confined on one brane and gravity is concentrated on the other. 
One extra dimension with a ``warped'' geometry is allowed to fluctuate, 
resulting in a massive scalar, the radion, in addition to massless 
(gravitons) and massive (accessible at future colliders) spin-two 
excitations. Gravity in the RS model is weak because it is exponentially 
damped by the distance between the branes. 
The radion has the same quantum numbers as the Higgs (but the radion 
can couple directly to gluons), and 
Higgs-radion mixing is possible. OPAL~\cite{op-ra} has looked at the 
``Higgs/radion-strahlung'' process, e$^+$e$^-$ $\rightarrow$ Z + h/r, 
reusing three searches for the Higgs boson. The excluded 
RS parameter space is obtained by a parameter scan (m$_r$, m$_h$, 
$\Lambda_W$ $\sim$ 1 TeV, mass scale on the SM brane, $\xi$, mixing 
parameter). For $\xi$ = 0 the SM Higgs mass limit is reobtained, for    
$\xi$ $\neq$ 0 the mass limit is generally lower and decreases  
with decreasing $\Lambda_W$. For all $\xi$, m$_r$, $\Lambda_W$ $\geq$ 246 
GeV, OPAL obtains m$_h$ $>$ 58 GeV at 95\% CL,~\cite{op-ra} the 
analyses losing their sensitivity for $\Lambda_W$ $\simeq$ 0.8 TeV.    
 
\section{Conclusions}

No evidence for single top production, 
excited leptons, or extra dimensions was  
found at LEP2 and exclusion limits were set. In many cases 
these results, four years after the LEP shut down,  
are still competitive with the new analyses produced 
at HERA and at the Tevatron.

\section*{Acknowledgments}
I would like to thank the colleagues who helped with the preparation 
of this talk: W. Adam, \,
S. Ask, L. Bellagamba, V. Hedberg, G. Paztor and P. Wells.

\end{document}